\documentclass[aps,prx,reprint,groupedaddress,superscriptaddress,footnoteinbib]{revtex4-2}
\usepackage{amsmath,amssymb}
\usepackage{bbold}
\usepackage{upgreek}
\usepackage{dsfont}
\usepackage{graphicx, epstopdf} 
\usepackage{amsfonts}
\usepackage{graphicx,xcolor}
\usepackage{amsmath, bbm}
\usepackage{natbib}
\usepackage{braket}
\usepackage[utf8x]{inputenc}
\usepackage[ngerman, english]{babel}
\usepackage[T1]{fontenc}
\usepackage{mathtools}
\usepackage{hyperref}
\usepackage{soul}

\begin{document}

\title{Amplifying a zeptonewton force with a single-ion nonlinear oscillator}

\author{Bo Deng}
\affiliation{Institute of Physics, University of Kassel, Heinrich-Plett-Stra{\ss}e 40, 34132 Kassel, Germany}

\author{Moritz G{\"o}b}
\affiliation{Institute of Physics, University of Kassel, Heinrich-Plett-Stra{\ss}e 40, 34132 Kassel, Germany}

\author{Benjamin A. Stickler}
\affiliation{Faculty of Physics, University of Duisburg-Essen, Lotharstra{\ss}e 1, 47057 Duisburg, Germany}

\author{Max Masuhr}
\affiliation{Institute of Physics, University of Kassel, Heinrich-Plett-Stra{\ss}e 40, 34132 Kassel, Germany}

\author{Kilian Singer}
\email{ks@uni-kassel.de}
\affiliation{Institute of Physics, University of Kassel, Heinrich-Plett-Stra{\ss}e 40, 34132 Kassel, Germany}

\author{Daqing Wang}
\email{daqing.wang@uni-kassel.de}
\affiliation{Institute of Physics, University of Kassel, Heinrich-Plett-Stra{\ss}e 40, 34132 Kassel, Germany}

\begin{abstract}
Nonlinear mechanical resonators display rich and complex dynamics and are important in many areas of fundamental and applied sciences. In this letter, we show that a particle confined in a funnel-shaped potential features a Duffing-type nonlinearity due to the coupling between its radial and axial motion. Employing an ion trap platform, we study the nonlinear oscillation, bifurcation and hysteresis of a single calcium ion driven by radiation pressure. Harnessing the bistability of this atomic oscillator, we demonstrate a 20-fold enhancement of the signal from a zeptonewton-magnitude harmonic force through the effect of vibrational resonance. Our findings open up a range of possibilities for controlling and exploiting nonlinear phenomena of mechanical oscillators close to the quantum regime.
\end{abstract}

\maketitle

Nonlinear oscillators, whose dynamics go beyond the simple description of linear equations of motion, are ubiquitous in nature. Driven nonlinear oscillators can exhibit complex behaviors such as squeezing\,\cite{Yang2021a, Huber2020}, persistent resonance\,\cite{Yang2021b}, phase transitions\,\cite{Stambaugh2006,Dolleman2019,Wang2020}, and chaos\,\cite{Testa1982,Lakshmanan1996Book}. They have a broad presence in many branches of science and technology, including signal processing, biomechanics, control systems, and force sensing \,\cite{Strogatz2018Book,Tadokoro2021,Cleland2002,Gieseler2013,Papariello2016,Zhang2020, Aldana2014}. Compared to linear oscillators, nonlinear oscillators can be more sensitive to external forces, especially when they are tuned close to a bifurcation point. The multistability of nonlinear oscillators provides opportunities for signal enhancement through the effects of stochastic and vibrational resonances\,\cite{Gammaitoni1998, Badzey2005,Almog2007,Landa2000,Chowdhury2020}. Both effects involve the introduction of an external modulation signal to the oscillator to amplify a weak signal under detection. Stochastic resonance\,\cite{Gammaitoni1998, Badzey2005, Almog2007} employs a broadband noise, whereas a high-frequency harmonic signal is used in vibrational resonance\,\cite{Landa2000, Chowdhury2020}.

The manifestation of these effects in mechanical systems has primarily been investigated with nano-electromechanical resonators\,\cite{Badzey2005,Almog2007,Chowdhury2020}. In this letter, we show that a particle trapped in a funnel-shaped three-dimensional potential exhibits an inherent Duffing-type nonlinearity. This nonlinearity arises from the coupling of radial and axial motion of the particle resulting from the geometry of the potential\,\cite{Rossnagel2016,Levy2020}. We investigate this model experimentally using a modified Paul trap that generates a funnel-shaped secular potential for a single calcium ion. Through radiation pressure forces, we actuate the radial motion of the ion and observe a bifurcation into two stable equilibrium states. By introducing an intermediate-frequency parametric modulation signal analogous to the case of vibrational resonance\,\cite{Landa2000, Chowdhury2020}, we show that a zeptonewton-scale oscillatory force acting on the axial mode can be enhanced by a factor of 20.

\begin{figure}[t]
\begin{center}
\includegraphics[width=1.0\columnwidth]{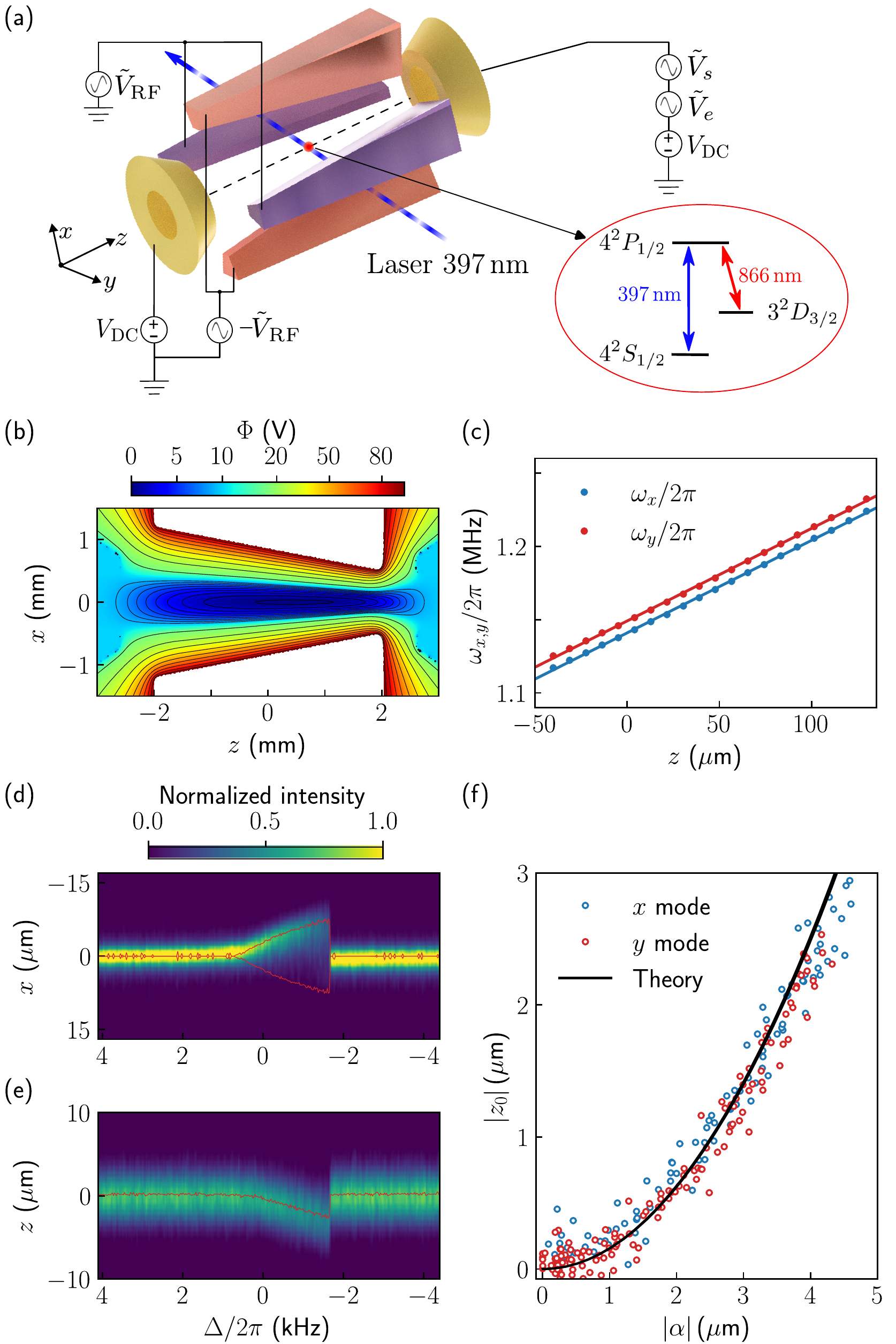}
\caption{(a) A laser-cooled ion confined in a funnel-shaped Paul trap. Intensity modulation of the laser excites the motion along the $x$ direction. Positive DC bias ($V_{\rm DC}$) and two AC voltages ({\small $\Tilde{V}_{\rm s}$}, {\small $\Tilde{V}_{\rm e}$}) are applied to the endcap electrodes. Inset: A simplified energy level diagram for the ${}^{40}\text{Ca}^{+}$ ion. The transition at 397\,nm is used for Doppler cooling, while the repumping laser at 866\,nm ensures a closed-cycle optical transition. (b) Electrostatic potential obtained from numerical simulations. (c) Dependence of radial trapping frequencies on the axial position. The dots represent measured data and the lines display linear fits. (d) Response of the $x$ mode to a descending frequency sweep. Each column represents the projection of a fluorescence image along the $z$ axis. The red lines display the extracted oscillation amplitude. (e) Response of the axial mode. The red line indicates the axial equilibrium position. (f) The amplitude of axial displacement plotted against the radial oscillation amplitude. The solid line shows the predictions following Eq.\,(\ref{eq:zeq}) in the absence of $F_z(t)$.}
\label{fig_funnel}
\end{center}
\end{figure}

As illustrated in Fig.\,\ref{fig_funnel}(a), our system consists of a single ion confined to the potential of a modified Paul trap. In contrast to conventional linear Paul traps, the two pairs of blade electrodes are inclined with respect to the axial $z$ axis\,\cite{Rossnagel2016,Levy2020}. We perform numerical field simulations using the boundary element method\,\cite{Betcke2021} to obtain the electrostatic potential resulting from this geometry. As displayed in Fig.\,\ref{fig_funnel}(b), this arrangement leads to a funnel-shaped potential in which the radial confinement becomes stronger with increasing axial position $z$. To trap a charged particle, the two pairs of blades are driven by bipolar radiofrequency signals {\small$\pm\Tilde{V}_{\rm RF}$}. This creates a secular trapping potential in the radial directions for a ${}^{40}\text{Ca}^{+}$ ion. The blue and red dots in Fig.\,\ref{fig_funnel}(c) display the measured secular trapping frequencies along the two radial principal axes $x$ and $y$ as a function of the axial position. Frequencies of both modes can be well recovered by the linear relation $\omega_{x,y} ( 1 + z/\ell_0)$, with $\omega_{x,y}\simeq2\pi\times\{1.14\,,1.15\}$\,MHz the trapping frequencies at $z = 0$ and $\ell_0=1.81 \,{\rm mm}$ a parameter describing the length of the funnel. In the axial direction, the ion is weakly confined by positive DC voltages (${V}_{\rm DC}$) applied on the two endcaps, resulting in an axial trapping frequency of $\Omega\simeq2\pi\times100$\,kHz.
 
The motion of the ion is subject to continuous damping in all three directions achieved through Doppler cooling using a laser red-detuned to its $4^2S_{1/2}-4^2P_{1/2}$ transition [see inset of Fig.\,\ref{fig_funnel}(a)]. A second laser at $866$\,nm repumps the metastable $3^2D_{3/2}$-state to facilitate a closed-cycle transition. In addition to the damping, the $x$- and $z$-motion are driven by external forces $F_x(t)$ and $F_z(t)$. Given that the frequency of the $y$-mode is sufficiently far from the frequencies of the other two modes, we consider that the $y$ mode is not driven and set $y = 0$ for all times. In the relevant regime of the experiment, the impact of the funnel shape is weak, $|z/\ell_0| \ll 1$. Keeping the $z$-dependence of the secular potential in linear order only, we obtain the Hamiltonian for the two-dimensional oscillation dynamics
\begin{align}\label{eq:hamiltonian}
H = & \frac{p_x^2 + p_z^2}{2m}+\frac{1}{2}m\Omega^2 z^2 + \frac{1}{2}m\omega_{x}^2\left(1+2\frac{z}{\ell_0}\right) x^2 \nonumber\\
& - F_x(t)x - F_z(t) z \,.
\end{align}

Here, $m$ represents the mass of the ${}^{40}\text{Ca}^{+}$ ion. The driving forces have three distinct contributions: (i) $F_x$ is realized by modulating the intensity of the cooling laser to produce a near-resonant radiation pressure force of magnitude $F_0$ and frequency $\omega_0 = \omega_x + \Delta \gg \Omega$ (with detuning $\Delta$, $|\Delta|\ll \Omega$), which drives the $x$-motion of the ion; (ii) a weak signal force of magnitude $F_{\rm s} $ and frequency $\omega_{\rm s} \ll \omega_0$ acting on the axial mode, which we want to detect; (iii) the axial enhancement force of strength $F_{\rm e}$ and frequency $\omega_{\rm e}$, which serves to enhance the detection of the signal force $F_{\rm s}$. In the experiment, the signal force is introduced by feeding in an AC voltage {\small$\Tilde{V}_{\rm s}$} with an amplitude of $500\,\mu{\rm V}$ on one endcap electrode. The detection scheme is not limited to electrostatic forces and should be applicable to any interaction that leads to an axial displacement of the ion. The enhancement force is applied in the same manner by exerting an AC voltage {\small$\Tilde{V}_{\rm e}$} on the same endcap electrode, as shown in Fig.\,\ref{fig_funnel}(a). In summary, $F_x(t) = F_0 \cos(\omega_0 t)$ and
\begin{equation}\label{eq:forcez}
    F_z(t) = F_{\rm e} \cos(\omega_{\rm e} t) + F_{\rm s} \cos(\omega_{\rm s} t) \,.
\end{equation}
Note that the frequencies of these three forces fulfill the relation $\omega_{\rm s} \ll \omega_{\rm e} \ll \omega_{\rm 0}$, as required for vibrational resonance\,\cite{Landa2000,Chowdhury2020}.

The equations of motion resulting from the Hamiltonian \eqref{eq:hamiltonian} can be simplified by exploiting the separation of timescales $\omega_0 \gg \Omega$. This is achieved by making the ansatz $x(t) = \alpha(t) e^{-i\omega_0 t} + {\rm c.c.}$ (with ${\rm c.c.}$ the complex conjugate), where the amplitude $\alpha$ evolves on a timescale much slower than $1/\omega_0$, so that $|\ddot \alpha| \ll \omega_0 |\dot{\alpha}|$ and $|\dot{\alpha}|\ll \omega_0 |\alpha|$. Averaging over one drive period $2\pi/\omega_0$ and performing a rotating-wave approximation yields the approximate equation of motion for the radial mode
\begin{align}\label{eq:effectivealpha}
    \dot{\alpha} = & i\left ( \Delta - \omega_x \frac{z}{\ell_0}\right )\alpha + i f_0  - \frac{\gamma}{2} \alpha \,,
\end{align}
with damping rate $\gamma$ and driving strength $f_0 = F_0/4 m \omega_x$. This equation shows that the axial displacement $z$ shifts the resonance frequency of the radial oscillator by $\omega_x z/\ell_0$.

The axial motion is described by
\begin{equation}
    \ddot{z} = - \Omega^2 z -  \frac{2 \omega^2_x}{\ell_0}|\alpha|^2 + \frac{F_z(t)}{m}- \gamma \dot{z} \,,
\end{equation}
which follows from the exact equations of motion via the substitution $x^2 = 2 |\alpha|^2$ due to the rotating-wave approximation. This equation implies that the instantaneous displacement of the axial equilibrium position 
\begin{equation}\label{eq:zeq}
    z_{0} = -2 \frac{\omega_x^2}{\Omega^2} \frac{|\alpha|^2}{\ell_0} + \frac{F_z(t)}{m \Omega^2}
\end{equation}
is determined by the radial oscillation amplitude as well as by the external force \eqref{eq:forcez}.

The coupling of radial and axial dynamics described by Eqs.\,\eqref{eq:effectivealpha} and \eqref{eq:zeq} is key to the emergence of nonlinearity in this system. We verify this coupling in our experiment by coherently driving the radial oscillator with the force $F_x$ and observing the response in the axial direction. To do so, the modulation frequency of the cooling laser $\omega_0$ is reduced stepwise and swept across $\omega_x$. In each step, a fluorescence image of the ion is recorded. The integration time of the camera is 400\,ms. Each frame therefore averages over many axial and radial oscillation cycles. The images are then projected along the radial and axial axes to separate the responses along these two directions. The magnitude of the radial force is limited by the saturation of the cooling transition to $\lesssim30$\,zN. The axial force $F_z$ is not applied during this measurement. Figure\,\ref{fig_funnel}(d) shows the response of the radial mode as a function of the frequency detuning $\Delta$. As $\omega_0$ approaches the resonance frequency, a coherent oscillation of the radial mode is excited. Here, only one flank of the oscillation is visible in the images due to a phase delay between the ion motion and the laser drive\,\cite{Drewsen2004}. The red lines in the graph display the extracted coherent oscillation amplitude $|\alpha|$. As the modulation frequency crosses $\omega_x$, the coherent oscillation persists for an extended range. The increase of radial oscillation amplitude is accompanied with an axial shift in the negative direction of the $z$ axis, i.e., towards the open end of the funnel, as displayed in Fig.\,\ref{fig_funnel}(e). The red line in this figure shows the extracted axial center position $z_0$. Figure\,\ref{fig_funnel}(f) displays a quantitative comparison of the measurement with Eq.\,\eqref{eq:zeq}. Blue and red circles represent the data measured for the $x$ and $y$ modes, which are in good match with the prediction of Eq.\,\eqref{eq:zeq} displayed by the black line.

At further frequency detuning, the oscillator experiences an abrupt change in both the radial and axial directions and resumes its initial position. Such an abrupt change is a signature of bistability in the system. Duffing-type bistability of trapped ions can arise from the intrinsic anharmonicity of linear Paul traps\,\cite{Akerman2010}. Here, the nonlinearity stems from the radial-axial coupling imposed by the funnel-shaped potential. This can be intuitively understood as shifting towards the open side of the funnel softens the radial spring and leads to an amplitude-dependent radial trapping frequency, i.e., a Duffing-type response\,\cite{Nayfeh2008book}.

To confirm the intuition, we solve the averaged radial amplitude equation\,\eqref{eq:effectivealpha}. For this, we use that the axial motion adiabatically follows the radial motion since $|\Delta|\ll \Omega$. One can thus average over one axial oscillation period $2\pi/\Omega$, replace $z$ by $z_0$ in Eq.~\eqref{eq:effectivealpha}, and obtain
\begin{equation}\label{eq:duffingalpha}
    \dot{\alpha} = i \left [ \Delta -\delta(t) \right ] \alpha + i \xi |\alpha |^2\alpha + i f_0 - \frac{\gamma}{2} \alpha \,.
\end{equation}
This equation describes a Duffing oscillator with a linear drive $f_0$ and a parametric drive that modulates the detuning by $\delta(t) =  \omega_x F_z(t)/m \ell_0\Omega^2$. Here, the parameter $\xi = 2 \omega_x^3/\Omega^2 \ell_0^2 $ quantifies the nonlinearity due to the radial-axial coupling. In our experiment, $\xi/2\pi = 9.04\times10^{13}\,\text{Hz}/\text{m}^{2}$ and significantly exceeds the intrinsic nonlinearity of the secular trapping potential.

\begin{figure}[t]
\includegraphics[scale=0.5]{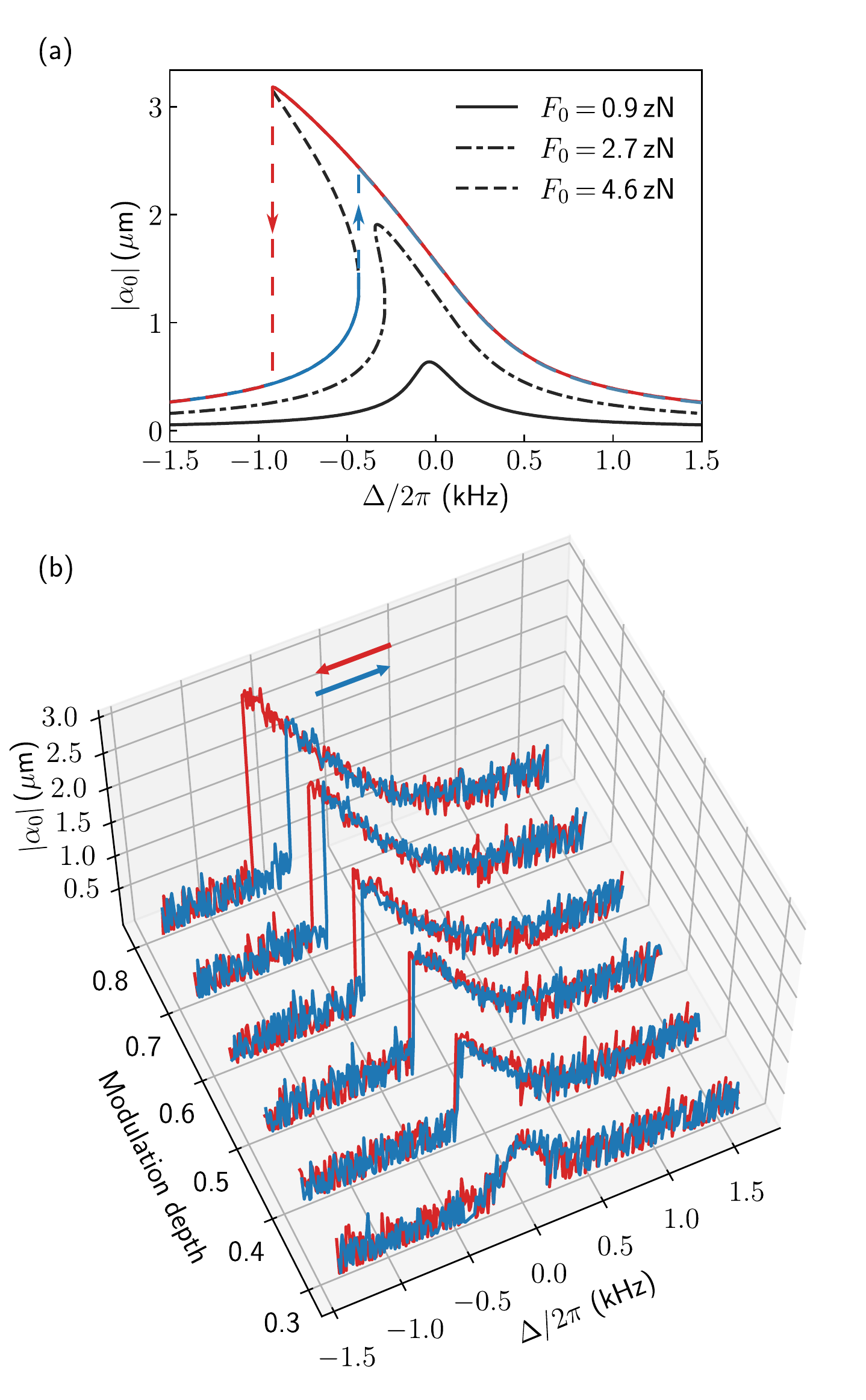}
\caption{(a) Solutions of the radial amplitude $|\alpha_0|$ for $\gamma=2\pi\times250$\,Hz and driving strengths denoted in the legend. (b) The emergence of bistability. The red lines show the radial amplitude when performing descending frequency sweeps and the blue lines represent the outcome for ascending frequency sweeps.}
\label{fig_hysteresis}
\end{figure}

Since the parametric driving frequencies [see Eq.~\eqref{eq:forcez}] are both much smaller than the damping rate, $\omega_{\rm s} \ll \omega_{\rm e} \ll \gamma$, the Duffing oscillator relaxes for each $\delta(t)$ towards a quasi-stationary state with the square of the amplitude $|\alpha_{0}(t)|^2$ determined by the third-order polynomial equation
\begin{equation}\label{eq:cubicsolution}
    |\alpha_{0}(t)|^2 = \frac{f_0^2}{[\Delta - \delta(t) + \xi |\alpha_{0}(t)|^2]^2 + \gamma^2/4} \,.
\end{equation}
The solutions of this equation as a function of the detuning $\Delta$ are displayed in Fig.\,\ref{fig_hysteresis}(a). Depending on the driving strength, it has either a single stable solution or three stationary solutions, of which two are stable and one is unstable\,\cite{Nayfeh2008book}. In the latter case, an adiabatic ramping of the detuning through the bistable region leads to hysteresis [see blue and red arrows in Fig.\,\ref{fig_hysteresis}(a)].

To verify the emergence of bistability, we measure the response of the radial oscillator to ascending and descending frequency sweeps. We gradually increase the modulation depth of the laser intensity, which increases the amplitude of the driving force. The results are displayed in Fig.\,\ref{fig_hysteresis}(b). Under weak driving, the same Lorentzian-shaped response is observed for ascending (blue lines) and descending (red lines) sweeps. As the driving force increases, the lineshape becomes unsymmetric and abrupt changes appear on the side of negative detuning in both ascending and descending measurements. With stronger driving, the jumps occur at different frequency detunings for descending and ascending sweeps, confirming that the oscillation amplitude has more than one stable solution in this regime. As predicted by theory, the hysteresis region opens gradually as the driving force further increases.

Having characterized our system, we continue to demonstrate that this nonlinear oscillator can be utilized for amplifying small signals through a scheme analogous to vibrational resonance. In typical vibrational resonance experiments\,\cite{Chowdhury2020}, a high-frequency amplitude modulation is introduced to a one-dimensional forced oscillator and enables switching between its two stable states. In our system, the three-dimensional feature allows us to explore a new scheme. Here, the radial mode is forced by $F_x$ to the bistable regime, and a small force $F_{\rm s}$ acting on the axial mode can be amplified by the axial enhancement force $F_{\rm e}$.

\begin{figure}[t]
\begin{center}
\includegraphics[width=1.0\columnwidth]{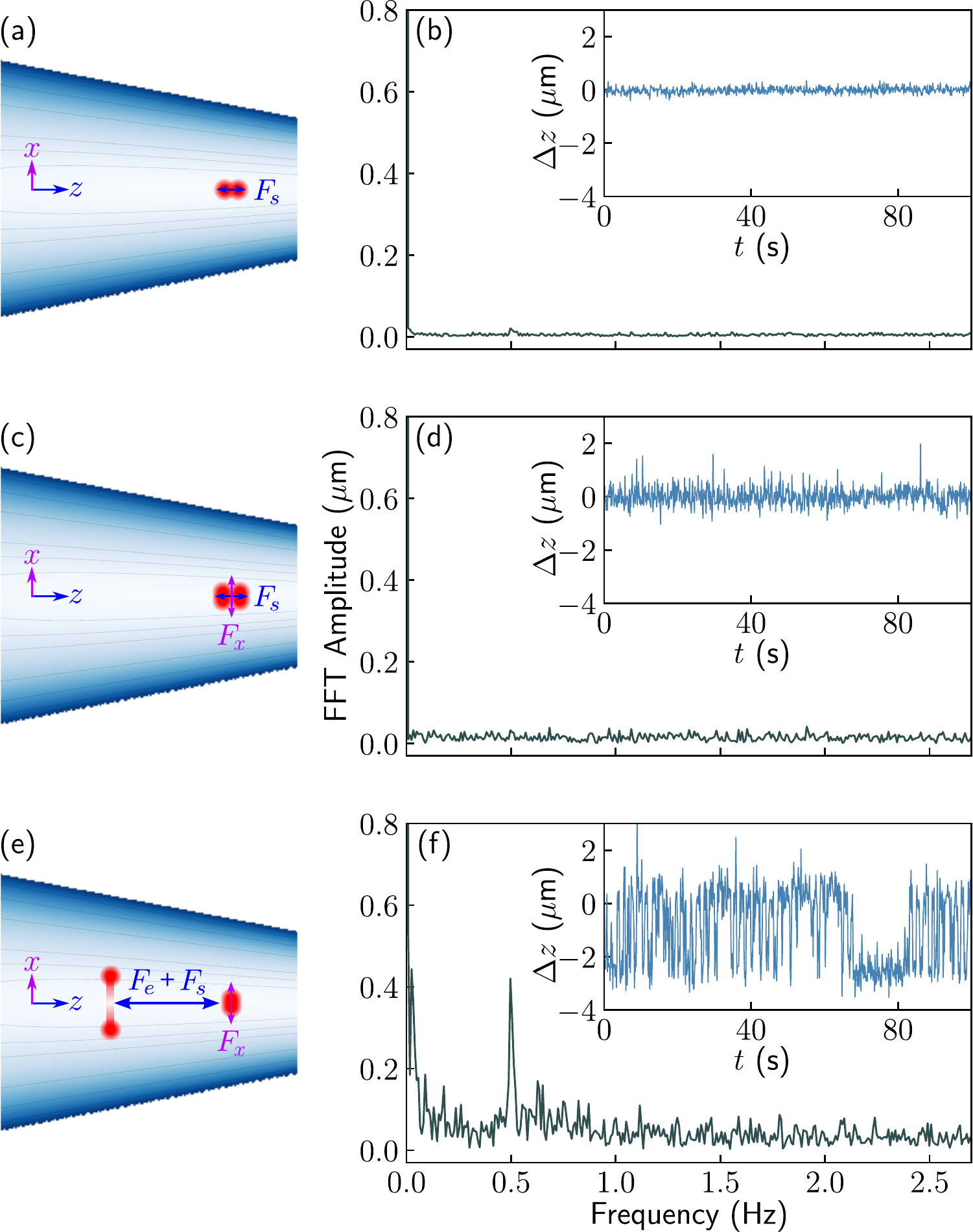}
\caption{Amplifying a weak signal force. (a) The weak signal force $F_s$ is applied on the axial oscillator ($F_x=0$ and $F_e=0$), introducing a small periodic axial displacement $\Delta z$ which is barely visible in the Fourier spectrum displayed in (b). The inset shows the extracted axial center position as a function of time. (c) The radial drive $F_x$ is applied. No enhancement is observed in the Fourier spectrum in (d). (e) The enhancement force $F_e$ is switched on. The trajectory in the inset of (f) shows clearly jumps between the two stable branches. The Fourier spectrum in (f) has a prominent peak at 0.5\,Hz.}
\label{fig_amplification}
\end{center}
\end{figure}

To start, we exert the signal force $F_{\rm s}$ by switching on the AC voltage {\small $\Tilde{V}_{\rm s}$} with an amplitude of 500\,$\mu{\rm V}$ at frequency $\omega_{\rm s}=2\,\pi\times0.5$\,Hz [see Fig.\,\ref{fig_funnel}(a) and the illustration in Fig.\,\ref{fig_amplification}(a)]. In this step, the radial forcing is not applied ($F_x=0$). The resulted signal force has a peak-to-peak amplitude of $2.4$\,zN and leads to a periodic displacement of $\pm45$\,nm of the axial oscillator. This displacement is comparable to the zero-point fluctuation of the axial mode, which amounts to $\sqrt{\hbar/2m\Omega}\approx36$\,nm. We determine the axial trajectory of the ion through super-resolution localization microscopy by performing a maximum likelihood estimation from the fluorescence images\,\cite{Lelek2021}. The resulted axial trajectory is displayed in the inset of Fig.\,\ref{fig_amplification}(b). Here, the camera is sampled at about 8\,Hz with an exposure time of 100\,ms for each frame. On average $N=240$ photons are detected per frame. The experimental axial Gaussian spread function has a standard deviation of $\sigma_0=1.64\,\mu{\rm m}$ and sets the lower bound of estimating the axial center position $\sigma_0/\sqrt{N}=106$\,nm\,\cite{Lelek2021}. Indeed, this small displacement is not directly resolvable in the trajectory. The harmonic nature makes it possible to identify this signal in the Fourier domain. As shown in Fig.\,\ref{fig_amplification}(b), a small peak at 0.5\,Hz is barely visible in the Fourier spectrum.

Next, we force the radial oscillator to the bistable regime by switching on the radial driving $F_x$. The frequency $\omega_0$ of the driving force is tuned to the center of the bistable region. The ion is initially prepared in the lower stable branch. The signal force $F_{\rm s}$ shifts the ion's motional state along the lower branch adiabatically but is too weak to induce a transition to the upper branch, as illustrated in Fig.\,\ref{fig_amplification}(c). The measured axial trajectory and the corresponding Fourier spectrum are displayed in Fig.\,\ref{fig_amplification}(d). The noise floor is elevated in the Fourier spectrum, but no enhancement of the signal at 0.5\,Hz is observed.

In the last step, we apply a sinusoidal enhancement signal {\small $\Tilde{V_{\rm e}}$} at $\omega_{\rm e}=2\pi\times50$\,Hz to one endcap, which exerts a parametric drive on the radial oscillator. The frequency of the enhancement force satisfies $\omega_{\rm s} \ll \omega_{\rm e} \ll \omega_{\rm 0}$ as required for vibrational resonance schemes\,\cite{Landa2000,Chowdhury2020}. By tuning the amplitude of {\small $\Tilde{V_{\rm e}}$} such that the parametric drive brings the oscillator to the boundaries of the bistable region, the signal force can introduce jumps of the oscillator to the upper branch and back [see Fig.\,\ref{fig_amplification}(e)]. The inset of Fig.\,\ref{fig_amplification}(f) shows the axial trajectory of the ion and clearly evidences the switch between two branches. In the Fourier spectrum, the strong peak at 0.5\,Hz lets us deduce an enhancement factor of 20 compared to the signals in (b) and (d). The maximal amplification that can be achieved is determined by the width of the hysteresis region. In our experiment, this is limited by the saturation of the cooling transition, which sets an upper bound to the radial driving force.

In summary, we have shown that a particle confined in a funnel-shaped potential exhibits a Duffing-type nonlinearity that originates from the coupling between its radial and axial motion. Using an ion trap platform, we observed the nonlinear oscillation, bifurcation and hysteresis of a laser-cooled ion oscillator in such a potential. Furthermore, we demonstrated that the bistability of this system can be exploited to enhance a zeptonewton-scale oscillatory force using a parametric variant of vibrational resonance. Our findings open up opportunities for combining the rich nonlinear dynamics with the precision of atomic physics to study and exploit nonlinear mechanical phenomena in single- and few-body systems close to their motional ground state. For instance, future work might assess the role of nonlinearity in energy transport through oscillator chains\,\cite{Zheng2001,Kovaleva2013,Borlenghi2015}, study synchronization of driven nonlinear oscillators in the quantum regime\,\cite{Walter2014,Lorch2016}, explore the emergence of limit cycles\,\cite{Arosh2021}, or investigate thermodynamics with quantum nonlinear oscillators\,\cite{Mendes2021}. The ability to enhance the detection of small forces could be extended to particles that are co-trapped with the atomic ion, such as molecules \cite{Wolf2016,Chou2017,Najafian2020}, and offers a promising route to detect forces that have not yet been observed on molecular systems, such as chiral optical forces\,\cite{Wolf2016,Chou2017,Najafian2020,Cameron2014,Canaguier-Durand2013,Stickler2021}. Last but not least, the geometry-induced nonlinearity could also be exploited to generate nonclassical states of massive particles that lack sharp internal transitions, such as electrically levitated nanoparticles\,\cite{Martinetz2020,Dania2021,Ren2022}.

\section*{Acknowledgement}
We thank Florian Elsen, David Zionski and Johannes Rossnagel for early-stage work on the ion trap and Klaus Hornberger for discussions on the theory. This work was supported by the Deutsche Forschungsgemeinschaft (DFG, German Research Foundation) – projects 499241080, 384846402, 328961117 – through the QuantERA grant ExTRaQT, the Research Unit Thermal Machines in the Quantum World (FOR 2724) and the Collaborative Research Center ELCH (SFB 1319).

\bibliographystyle{apsrev4-2}
\end{document}